\documentclass[aps,prl,reprint,superscriptaddress,showpacs,10pt,tightenlines]{revtex4-1}
\usepackage{graphicx}
\usepackage{dcolumn}
\usepackage{bm}
\usepackage{amsmath}
\usepackage[dvipsnames]{xcolor}
\usepackage{amssymb}
\usepackage{amsthm}
\usepackage[titletoc]{appendix}
\usepackage{hyperref}
\usepackage[utf8]{inputenc}
\usepackage[english]{babel}
\newtheorem{theorem}{Theorem}
\newtheorem{prop}{Proposition}
\newtheorem{lemma}{Lemma}
\newtheorem{corollary}{Corollary}
\newcommand{\WH}{\widetilde{H}}
\newcommand{\pk}{\psi_{k}}
\newcommand{\wpk}{\widetilde{\psi}_{k}}

\newcommand{\wz}{\widetilde{0}}
\newcommand{\wo}{\widetilde{1}}
\newcommand{\wu}{\widetilde{U}}
\hypersetup{colorlinks=true, citecolor=blue, urlcolor=blue, linkcolor=blue}
\begin{document}
\title{Spectral Analysis of Product Formulas for Quantum Simulation}
\author{Changhao Yi}
\email{yichanghao123@unm.edu}
\affiliation{Center for Quantum Information and Control, University of New Mexico}
\author{Elizabeth Crosson}
\email{crosson@unm.edu}
\affiliation{Center for Quantum Information and Control, University of New Mexico}
\date{\today}

\begin{abstract}
We consider Hamiltonian simulation using the first order Lie-Trotter product formula under the assumption that the initial state has a high overlap with an energy eigenstate, or a collection of eigenstates in a narrow energy band.  This assumption is motivated by quantum phase estimation (QPE) and digital adiabatic simulation (DAS).   Treating the effective Hamiltonian that generates the Trotterized time evolution using rigorous perturbative methods, we show that the Trotter step size needed to estimate an energy eigenvalue within precision $\epsilon$ using QPE can be improved in scaling from $\epsilon$ to $\epsilon^{1/2}$ for a large class of systems (including any Hamiltonian which can be decomposed as a sum of local terms or commuting layers that each have real-valued matrix elements).  For DAS we improve the asymptotic scaling of the Trotter error with the total number of gates $M$ from $\mathcal{O}(M^{-1})$ to $\mathcal{O}(M^{-2})$, and for any fixed circuit depth we calculate an approximately optimal step size that balances the error contributions from Trotterization and the adiabatic approximation.  These results partially generalize to diabatic processes, which remain in a narrow energy band separated from the rest of the spectrum by a gap, thereby contributing to the explanation of the observed similarities between the quantum approximate optimization algorithm and diabatic quantum annealing at small system sizes.   Our analysis depends on the perturbation of eigenvectors as well as eigenvalues, and on quantifying the error using state fidelity (instead of the matrix norm of the difference of unitaries which is sensitive to an overall global phase).
\end{abstract}
\maketitle

\section{Introduction}

The Lie-Trotter product formula~\cite{trotter1959product,suzuki1976generalized} was originally used by Lloyd~\cite{lloyd1996universal} to establish the first method for efficiently approximating the dynamics $U(t) = e^{-i t H}$ generated by a local Hamiltonian $H$ with a universal quantum computer. After many refinements~\cite{childs2019theory, childs2019nearly, campbell2019random} this approach (often called ``Trotterization'') continues to be an appealing method for Hamiltonian simulation from both experimental and mathematical perspectives. 

The method is based on dividing $U(t)$ into $L$ short-time evolutions $U(t) = U^{L}(\delta t), t = L \delta t$, and replacing each $U(\delta t)$ with an approximation $T(\delta t)$.  The parameter $L$ is the number of Trotter steps and $\delta t > 0$ is the Trotter step size.     Given a decomposition of the Hamiltonian into a sum of layers $H = \sum_{n=1}^\Gamma H_n$~\footnote{As usual in trotterization, the guideline for the decomposition is that one has a way to implement $e^{i t H_j}$ efficiently for each $H_j$.   A sufficient condition for this is for each $H_j$ to be a sum of pairwise commuting local terms.}, the first order product formula approximation is
\begin{equation}
T(\delta t) \equiv \prod_{n=1}^{\Gamma}e^{-iH_{n}\delta t},
\label{equ:1st}
\end{equation}
where $\delta t$ and $L$ are chosen to depend on the tolerable level of error, as we subsequently discuss.

Most prior works quantify the Trotter error in terms of the operator norm $\|U(\delta t) - T(\delta t)\|$, but in this work we directly compare states evolved under the exact and approximate time evolution to produce a tighter error estimate for the specific class of initial states we consider.    We quantify the Trotter error in terms of the phase error $\theta$ and the fidelity error $f$,
\begin{equation}
T(\delta t)^{L}|\psi\rangle = \sqrt{1-f}e^{i\theta}U(t)|\psi\rangle + \sqrt{f}U(t)|\psi^{\perp}\rangle.
\label{equ:setup}
\end{equation}
where $|\psi\rangle$ is the initial state.  While the operator norm can always be used to upper bound $f$ and $\theta$ for an arbitrary initial state $|\psi\rangle$, we identify situations in which analyzing $f$ and $\theta$ for specific initial states produces tighter error bounds than the general case.  

Our results are based on a spectral analysis of the effective Hamiltonian $\WH$ that generates the Trotterized time evolution, 
\begin{equation}
\WH \equiv i\log(T(\delta t))/\delta t,\quad T(\delta t) = e^{-i\WH \delta t}
\label{equ:eff_H}
\end{equation}
Regarding $\delta t$ as a small parameter, perturbative methods can be used to compare the spectrum $\{E_k ,|\psi_k\rangle\}$ of the original Hamiltonian $H$ to the spectrum $\{\widetilde{E}_k ,|\widetilde{\psi}_k\rangle\}$ of the effective Hamiltonian $\tilde{H}$.  These methods lead us to consider applications in which the initial state $|\psi\rangle$ is (or is close to) an eigenstate of $H$, enabling an improved upper bound on the Trotter step size in Quantum Phase Estimation (QPE) and Digital  Adiabatic Simulation (DAS).

QPE is one of the most important quantum simulation algorithms~\cite{kitaev1995quantum}, which relates Hamiltonian time-evolution $U(t) = e^{-i t H}$ to the measurement of energy eigenvalues~\cite{reiher2017elucidating,kivlichan2020improved, nielsen2002quantum}. In the ideal version of the algorithm, measuring the output of the phase estimation circuit collapses the system into an energy eigenstate of $H$. If we replace the time evolution with the product formula approximation $T(\delta t)^{L}$, then (in the ideal case) we will instead measure an energy eigenvalue of $\widetilde{H}$.   Therefore the Trotter error is directly related to the difference in spectrum between $H$ and $\WH$. Under conditions which are satisfied in many cases of interest, we show that the first order perturbative correction vanishes and use this to rigorously show that the Trotter error in phase can be reduced from $\mathcal{O}(L\delta t^{2})$ to $\mathcal{O}(L\delta t^{3})$.  In terms of the target precision $\epsilon$ of the QPE, this means the Trotter step size can be enlarged from $\delta t = \mathcal{O}(\epsilon)$ to $\delta t = \mathcal{O}(\epsilon^{1/2})$.  

DAS can be used to implement adiabatic quantum computation \cite{farhi2001quantum,aharonov2008adiabatic,albash2018adiabatic} and adiabatic state preparation~\cite{aharonov2003adiabatic, wan2020fast} on a digital quantum computer.  The scaling of Trotter error in DAS has previously been analyzed in terms of the operator norm~\cite{barends2016digitized, sun2018adiabatic}.  We find this measure of error is dominated by the accumulation of a global phase, whereas in adiabatic algorithms it is generally only the fidelity error $f$ that matters.  We regard the Trotterized time evolution as a discretized adiabatic evolution under an effective Hamiltonian, and obtain tighter bounds on the fidelity error by applying an adiabatic theorem to the effective Hamiltonian.  

This paper is organized as follows: we first illustrate the setting of the problems and our main results, together with the techniques and lemmas used in the proof. The primary mathematical tools are a rigorous perturbation method~\cite{ambainis2004elementary} and the Magnus expansion~\cite{blanes2009magnus}. Then we apply the main results to analyze Trotter error in QPE and DAS, in both cases we achieve improvements in circuit complexity. 


\section{Main Results}

\subsection{Set up and Notations}

Usually, the Trotter error is quantified by the norm distance between operators:
\begin{equation}
\Delta \equiv \|\hat{\Delta}\|,\quad\hat{\Delta} \equiv T(\delta t)^{L} - U(t)
\end{equation}
The notation $\|\cdot\|$ refers to the operator norm : $\|M\| = \max_{\|x\|_{2} = 1}\|M|x\rangle\|_{2}$, where $\|\cdot\|_{2}$ is the Euclidean norm of vector $\|v\|_{2} = \sqrt{vv^{\dagger}}$.  To quantify $\Delta$, it's enough to quantify the norm distance error of a single Trotter step $\delta \equiv \|T(\delta t) - U(\delta t)\|$ as $\Delta  \le L\delta$. For a given error tolerance $\epsilon$, the restriction of $\Delta \le \epsilon$ determines the gate complexity of the algorithm. 

In this paper, we separate the digital error into phase error $\theta$ and fidelity error $f$ defined by
\begin{gather}
f \equiv 1 - |\langle\psi|U^{\dagger}(t)T(\delta t)^{L}|\psi\rangle|^{2}\\
\theta \equiv \text{Arg}\left(\langle\psi|U^{\dagger}(t)T(\delta t)^{L}|\psi\rangle\right)
\end{gather}
where $|\psi\rangle$ is an initial state. For any $L,\delta t$ that satisfy $\Delta \le 1/\sqrt{2}$, the Euclidean distance error $\mathcal{E} \equiv \|\hat{\Delta}|\psi\rangle\|_{2}$ satisfies (see Appendix A)
\begin{equation}
f + \frac{\theta^{2}}{4} \le \mathcal{E}^{2} \le 2f + \theta^{2}
\label{equ:fandtheta}
\end{equation}
Without further assumptions about $H$ and $|\psi\rangle$, the two parameters are bounded by $f = \mathcal{O}(\Delta^{2}), |\theta| = \mathcal{O}(\Delta)$ as $\mathcal{E} \le \Delta$.  However, in some special cases, we find that both $f$ and $\theta^{2}$  have a different parameteric scaling with $\Delta$. This means that $\Delta$ does not always reflect the true Trotter error that we are interested in, which leads us to a different approach.

\subsection{Improved $f$ : Spectral Analysis}

The Trotterized evolution operator $T(\delta t)^{L}$ can be viewed as an exact evolution under an effective Hamiltonian $\WH \equiv i\log(T(\delta t))/\delta t$. Owing to the tiny size of $\delta t$, the spectrum of $\widetilde{H}$ is $\{\widetilde{E}_{k},|\wpk\rangle\}$ is close to that of $H$. Based on this observation,
suppose the initial state is one of the eigenstate $|\psi_{k}\rangle$ of $H$, $f$ and $\theta$ can be quantified by the difference between spectrum:
\begin{gather}
f = \mathcal{O}(1-|\langle \pk|\wpk\rangle|^{2})\\
\theta = \mathcal{O}(|\widetilde{E}_{k} - E_{k}|t)
\label{equ:relation}
\end{gather}
To ensure there exists a one-to-one correspondence between $|\psi_k\rangle$ and $|\tilde{\psi}_k\rangle$, we assume the spectrum is nondegenerate.  Therefore there is some spectral gap $\lambda_k = \min(E_k - E_{k-1}, E_{k+1}-E_k)$ around this eigenstate.   We use $\lambda$ as a general lower bound for the spectral gap around an initial eigenstate.   

The upper bound for the difference between energy eigenvalues is based on an upper bound for $\|\WH\|$. Using the Baker-Campbell-Hausdorff formula, 
\begin{equation}
\widetilde{H}(\delta t) = H + \frac{i\delta t}{2}\sum_{l>m}[H_{l},H_{m}] + \mathcal{O}(\delta t^{2}).
\label{equ:linear}
\end{equation}
The first few terms of the standard (Rayleigh-Schrodinger) perturbation theory can be used to estimate $f$ and $\theta$, but to avoid convergence issues and derive rigorous results we use other methods ~\cite{ambainis2004elementary, jansen2007bounds} that are widely used in proofs of adiabatic theorems. By Weyl's inequality, the perturbation in the eigenvalues satisfies
\begin{equation}
|\widetilde{E}_{k} - E_{k}| \le \|\WH - H\| \\
\end{equation}
The perturbation of the eigenvectors is derived from the following lemma.

\begin{lemma} [Rigorous perturbation method~\cite{jansen2007bounds}] $H(s)$ is a parameterized Hamiltonian with spectrum :$\{E_{j}(s), P_{j}(s)\}$, define
\begin{equation*}
P(s) = \sum_{j=1}^{m}P_{j}(s)
\end{equation*}
as the projector into a subspace $\mathcal{A}$ spanned by $m$ eigenstates. Its derivative has norm upper bound:
\begin{equation}
\|P^{\prime}(s)\| \le \sqrt{m}\|H^{\prime}(s)\|/\lambda
\end{equation} 
where $\lambda$ is the lower bound of energy gap between the eigenstates in and outside region $\mathcal{A}$.
\label{lemma : rigorp}
\end{lemma}
For a single eigenvector, lemma \ref{lemma : rigorp} implies
\begin{equation}
\begin{aligned}
\sqrt{1 - |\langle \wpk|\pk\rangle|^{2}} &= \|\widetilde{P}_{k} - P_{k}\|\\
&= \|P(\delta t) - P(0)\| \le \delta t\max_{s\in[0,\delta t]}\|P^{\prime}(s)\|\\
&\le \max_{s\in[0,\delta t]}\|\WH^{\prime}(s)\|\delta t/\lambda
\label{equ:norm_to_fid}
\end{aligned}
\end{equation}
The term $\|\widetilde{H}'(s)\|$ quantifies the size of the perturbation.   The necessary bound on $\|\WH^{\prime}\|$ has already been obtained using the Magnus expansion in previous work, see Appendix B for more details.

\begin{lemma}[Magnus Expansion\cite{tran2020faster}] Given $\widetilde{H}$ defined in Eq : (\ref{equ:eff_H}), when $\delta t = \mathcal{O}(N^{-1})$, for all $s\in [0,\delta t]$:
\begin{gather}
\|\WH(s) - H\| = \mathcal{O}(h\delta t),\quad \|\WH^{\prime}(s)\| = \mathcal{O}(h)\\
h = \frac{\alpha}{2} + \frac{4}{3}(\beta + 128\alpha\|H\|)\delta t\\
\alpha \equiv \sum_{n>m}||[H_{n},H_{m}]||\nonumber \\
\beta \equiv \sum_{l\ge n>m}||[H_{l},[H_{n},H_{m}]]|| \nonumber
\end{gather}
\label{lemma:magnus}
\end{lemma}
For example, if $H = \sum_{j = 1}^N h_j$ is a local Hamiltonian on $N$ qubits that satisfies $[h_{j},h_{k}] = 0, \forall |j-k|>1$, then $h = \mathcal{O}(N) + \mathcal{O}(N^{2}\delta t) = \mathcal{O}(N)$. In general, the parameter dependence of $h$ is complicated but will always be $\textrm{poly}(N)$ for any $k$-local Hamiltonian. To simplify the notation we retain the form of $h$ in the following paragraph.  

The following result follows from Lemma \ref{lemma : rigorp} with Lemma \ref{lemma:magnus}, see Appendix C. 
\begin{corollary}[Eigenstate as initial state]
For any $H$ and any eigenstate $|\pk\rangle$ of $H$ separated from the rest of the spectrum by a spectral gap $\lambda$, the time evolution under the 1st-order product formula satisfies :
\begin{equation*}
T(\delta t)^{L}|\pk\rangle = \sqrt{1-f}e^{i\theta} U(t)|\pk\rangle + \sqrt{f}|\pk^{\perp}\rangle
\end{equation*}
where the fidelity error and phase error satisfy
\begin{gather}
|\theta| = \mathcal{O}(Lh\delta t^{2})\\
f = \min\left\{\mathcal{O}\left(\frac{h^{2}\delta t^{2}}{\lambda^{2}}\right), \mathcal{O}(L^{2}h^{2}\delta t^{4})\right\}
\end{gather}
with $h$ defined in Lemma \ref{lemma:magnus}.
\label{coro:eigen}
\end{corollary}
As a comparison,  $\Delta = \mathcal{O}(Lh\delta t^{2})$ in general. Corollary \ref{coro:eigen} implies that the fidelity error $f$ eventually stops growing with with the total number of Trotter steps $L$, which is an extreme example of $f \ll \Delta^{2}$. After a short initial period, the Trotter error only accumulates in the global phase. This fact can be related to the leakage rate property~\cite{csahinouglu2020hamiltonian} of the Trotterized evolution operator. Suppose the initial state is $|\psi\rangle = \sum_{k}c_{k}|\psi_{k}\rangle$, where $|\psi_{k}\rangle$ all belong to a special region $\mathcal{A}$, like low-energy states. The leakage rate is the percentage for $|\psi\rangle$ to go outside that region after $T(\delta t)^{L}$: $1 - \text{Tr}(P_{\mathcal{A}}T(\delta t)^{L}|\psi\rangle\langle\psi|T^{\dagger}(\delta t)^{L})$. Using the argument about $\WH$ we prove:
\begin{theorem} [Leakage rate]
$T(\delta t)^{L}$ is Trotterized evolution operator, if the initial state $\rho$ belongs to subregion $\mathcal{A}$ spanned by $m$ eigenstates:
\begin{equation*}
P = \sum_{j=1}^{m}P_{j},\quad \text{Tr}(\rho P) = 1
\end{equation*} 
Then the leakage rate can be bounded by the norm distance between $P$ and corresponding effective projector $\widetilde{P}$ induced by $\WH$ if $\|P - \widetilde{P}\| < 1$:
\begin{equation*}
\begin{aligned}
1 - \text{Tr}(T(\delta t)^{L}\rho T^{\dagger}(\delta t)^{L}P) &= \mathcal{O}\left(\|P - \widetilde{P}\|^{2}\right)\\
&= \mathcal{O}\left(\frac{mh^{2}\delta t^{2}}{\lambda^{2}}\right)
\end{aligned}
\end{equation*}
where $\lambda$ is the lower bound of energy gap between the eigenstates in and outside region $\mathcal{A}$.
\label{coro:multi}
\end{theorem} 

\subsection{Improved $\theta$ : Special Perturbation}

The bound $\theta = \mathcal{O}(\Delta)$ can be tight, even when the initial state is an eigenstate. However, we show that it can be improved by a factor of $\delta t$ under assumptions that are satisfied for many Hamiltonians of interest, and this improves the scaling of the Trotter step size needed for QPE.    This can be illustrated by the simple case $H = H_A + H_B$ in which the Hamiltonian is decomposed into two commuting layers,
\begin{gather*}
T(\delta t) = e^{-iH_{A}\delta t}e^{-iH_{B}\delta t} = e^{-i\widetilde{H}\delta t},\\
\widetilde{H} = H - i\frac{\delta t}{2}[H_{A},H_{B}] + \mathcal{O}(\delta t^{2}).
\end{gather*}
The leading perturbation term is $V \equiv -i\delta t[H_{A},H_{B}]/2$. In standard perturbation theory, the 1st order correction in energy is $E^{(1)} = \langle \pk|V|\pk\rangle$. However, 
\begin{equation*}
\langle \psi_{k}|[H_{A},H_{B}]|\psi_{k}\rangle = \langle \psi_{k}
|[H,H_{B}]|\psi_{k}\rangle = 0.
\end{equation*}
In previous section we prove an upper bound of Trotter error in energy of order $\mathcal{O}(h\delta t)$. While under this special situation, because $E^{(1)} = 0$, the shift in energy at most has order $\mathcal{O}(\delta t^{2})$. This improvement can be applied to a general decomposition $H = \sum_{n=1}^{\Gamma}H_{n}$.   Whenever the leading order correction,
\begin{equation*}
V = \frac{i\delta t}{2}\sum_{l>m}[H_{l},H_{m}],
\end{equation*}
is off-diagonal in the eigenbasis of $H$, we can reduce the Trotter error in energy from $\mathcal{O}(\delta t)$ to $\mathcal{O}(\delta t^{2})$.  In the setting of Corollary \ref{coro:eigen}, the following result is proven by rigorous perturbative methods in Appendix E. 
\begin{lemma}$H$ is a normalized local Hamiltonian on $N$ sites with spectrum $\{E_{k},|\psi_{k}\rangle\}$, $\WH$ is its corresponding effective Hamiltonian induced from 1st order product formula. The new spectrum is $\{\widetilde{E}_{k},|\widetilde{\psi}_{k}\rangle\}$. The first perturbation of $\WH$ is off-diagonal in the eigenbasis of $H$:
\begin{equation}
\forall |\pk\rangle,\quad \langle\pk|\WH - H|\pk\rangle = \mathcal{O}(\delta t^{2})
\label{equ : criteria}
\end{equation} 
Further if $\delta t = \mathcal{O}(\lambda/N)$, where $\lambda$ is the lower bound of spectral gap between $|\psi_{k}\rangle$ and neighboring eigenstates, then the shift in energy satisfies: 
\begin{equation}
|\widetilde{E}_{k} - E_{k}| = \mathcal{O}\left(N^{2}\delta t^{2}\max\left\{1,\frac{1}{\lambda^{2}}\right\}\right)
\end{equation}
\label{lemma:spe}
\end{lemma}
In addition to Hamiltonians that can be decomposed into two commuting layers (of which a prominent class of examples are Hamiltonians $H = H_X + H_Z$ that have local terms which involve only Pauli $X$ or Pauli $Z$ operators), we list other different conditions where Eq : (\ref{equ : criteria}) is satisfied. 
\begin{itemize}
\item Real Hamiltonians.  Assume all of the local terms of $H$ have real matrix elements in some basis.  The components of an eigenstate $|\pk\rangle$ of any real symmetric matrices can all be taken to be real.  Consider an arbitrary commutator in $V$, $\langle\pk|H_{l}H_{m}|\pk\rangle$ is conjugate to $\langle\pk|H_{m}H_{l}|\pk\rangle$, and both are real numbers. So they are equal and appear with opposite signs in the commutator. Therefore, $\forall k,l,m, \langle\pk|[H_{l},H_{m}]|\pk\rangle = 0$.

\item Any Hamiltonian whose layers (or local terms) can be totally ordered to satisfy $[H_{l},H_{m}] = 0, \forall |l-m|>1$.  This condition is satisfied by 1D Hamiltonians with nearest-neighbor interactions, as well as general lattice Hamiltonians regarded as 1D chains of super-sites (since our results do not depend on the local dimension).   This condition leads to a recursive relation: $[H_{l},H_{l+2}] = [H_{l}, H - H_{l-1} - H_{l+1}] = 0$. If $[H_{l-1},H_{l}]$ is off-diagonal, $[H_{l},H_{l+1}]$ is also off-diagonal as any operator in the form of $[O,H]$ is off-diagonal in the eigenbasis of $H$. The case of $l=1$ is special as we don't have $H_{0}$. Thus $[H_{1}, H_{2}] = [H_{1},H]$ is off-diagonal. As a result, $\forall l, [H_{l},H_{l+1}]$ is off-diagonal.  These assumptions can be satisfied for any Hamiltonian with geometrically local terms in 1D.  
 
\item Frustration-free Hamiltonians~\cite{bravyi2010complexity}. This type of Hamiltonian satisfies $H = \sum_{j}\Pi_{j}, H|\psi_{0}\rangle = E|\psi_{0}\rangle, \Pi_{j}|\psi_{0}\rangle = E_{j}|\psi_{0}\rangle$ where $|\psi_{0}\rangle$ is ground state. With this property, when the initial state is the ground state, there will be no Trotter error no matter how big $\delta t$ is. In Lemma \ref{lemma:spe}, the upper bound on the Trotter error is inversely proportional to the spectral gap $\lambda$. However, it's possible for frustration-free Hamiltonian to be gapless~\cite{bravyi2012criticality}. This example shows that our methods can still overestimate the Trotter error for gapless Hamiltonians.

\end{itemize}
The second part of Eq : (\ref{equ:fandtheta}) indicates that $f$ and $\theta$ can be used to bound the Euclidean distance error as well. We have just proved that $f$ and $\theta^{2}$ can both be much smaller than $\Delta$, which means under these conditions the norm distance can't reflect Trotter error in Euclidean distance either.
 
Finally, we provide an example $H = H_1 + H_2 + H_3$ in which $V$ is not off-diagonal to show the result in Lemma \ref{lemma:spe} is not fully general. Let $H$ be a diagonal matrix $\Lambda$ in the eigenbasis of itself. In this basis choose:
\begin{equation*}
H_{1} = X\otimes I,\quad H_{2} = Y\otimes I,\quad H_{3} = \Lambda - H_{1} - H_{2}
\end{equation*}
Thus:
\begin{equation*}
V/\delta t = i[(X+Y)\otimes I,\Lambda]/2 + Z\otimes I
\end{equation*}
The first term is off-diagonal, the second term is not. Thus $V$ is not off-diagonal.

\subsection{Improved $f$ : Adiabatic Theorem}

DAS is a special type of time-dependent evolution simulation task~\cite{poulin2011quantum,berry2020time} that leverages the quantum adiabatic theorem~\cite{cheung2011improved,jansen2007bounds}. Physically, when the Hamiltonian evolves with time slowly enough, an initial state in some eigenspace will stay close to that eigenspace of the time-dependent Hamiltonian at all times. The evolution operator under $\hat{H}(t)$ has expression:
\begin{gather*}
i\frac{d}{dt}|\psi(t)\rangle = \hat{H}(t)|\psi(t)\rangle,\quad |\psi(T)\rangle = \hat{A}(T)|\psi(0)\rangle\\
\hat{A}(T) = \exp_{\mathcal{T}}\left(-i\int_{0}^{T}\hat{H}(t)dt\right)
\end{gather*}
We restrict out attention to the linear adiabatic path $\hat{H}(t) = (1-t/T)H_{i} + t/T H_{f}$.   In terms of the dimensionless parameter $s = t/T$, 
\begin{gather*}
H(s) \equiv \hat{H}(Ts) = (1-s)H_{i} + sH_{f}\\
A(T) \equiv \exp_{\mathcal{T}}\left(-iT\int_{0}^{1}H(s)ds\right) = \hat{A}(T)
\end{gather*}
$T\to\infty$ corresponds to the case where evolution is performed adiabatically:
\begin{gather*}
\lim_{T\to \infty}A(T)|\psi_{i}\rangle = |\psi_{f}\rangle
\end{gather*}
$|\psi_{i}\rangle$ is one eigenstate of initial Hamiltonian $H_{i}$ and $|\psi_{f}\rangle$ is the corresponding one of $H_{f}$. $T$ quantifies how slowly the evolution is, it also reflects how close the evolved state is to $|\psi_{f}\rangle$.  We apply the following rigorous adiabatic theorem.
\begin{lemma}[Adiabatic theorem~\cite{jansen2007bounds}] The error of adiabatic evolution is quantified by:
\begin{equation}
\epsilon_{adb} \equiv \|P_{f} - A(T)P_{i}A^{\dagger}(T)\|
\end{equation}
where $P_{f} = |\psi_{f}\rangle\langle\psi_{f}|$, $P_{i} = |\psi_{i}\rangle\langle\psi_{i}|$. $A(T)$ is the adiabatic evolution operator. Then:
\begin{equation}
\epsilon_{\text{adb}} \le \mathcal{G}(T,H)
\end{equation}
with
\begin{equation}
\begin{aligned}
\mathcal{G}(T,H) &\equiv \frac{1}{T}\left(\frac{\|H^{\prime}(0)\|}{\lambda(0)^{2}} + \frac{\|H^{\prime}(1)\|}{\lambda(1)^{2}}\right)\\
& + \frac{1}{T}\int_{0}^{1}\frac{\|H^{\prime\prime}(s)\|}{\lambda^{2}(s)} + 7\frac{\|H^{\prime}(s)\|^{2}}{\lambda^{3}(s)}ds
\end{aligned}
\end{equation}
$\lambda(s)$ is the lower bound of energy gap between $|\psi(s)\rangle$ and other eigenstates during evolution. $|\psi(s)\rangle$ is the eigenstate in $H(s)$ associated with $|\psi_{i}\rangle$. 
\label{lemma:continu}
\end{lemma}
\noindent Notice the adiabatic theorem refers to fidelity error,
\begin{equation*}
\epsilon_{\text{adb}} = \sqrt{1 - |\langle\psi_{f}|A(T)|\psi_{i}\rangle|^{2}}.
\end{equation*}
To simulate this process with Trotterization, we first approximate $A$ by a product of short time evolutions,
\begin{gather}
A_{d} \equiv \prod_{a=1}^{M}U_{a},\quad U_{a} \equiv e^{-iH_{a}\bar{\delta t}}\\
\bar{\delta t} \equiv \frac{T}{M},\quad H_{a} \equiv H\left(\frac{a}{M}\right)
\end{gather}
Here $M$ is the discretization number.  Note that going from $A$ to $A_d$ already incurs some discretization error.   Each short time evolution operator is further Trotterized,
\begin{gather}
A_{t} \equiv \prod_{a=1}^{M}U^{t}_{a},\quad U^{t}_{a} \equiv U^{t}\left(\frac{a}{M}\right)\\
U^{t}(s) \equiv e^{-i\bar{\delta t}(1-s)H_{i}}e^{-i\bar{\delta t} s H_{f}}
\label{equ:dig_adb}
\end{gather}
Therefore the parameter $M$ determines both the discretization error and the additional Trotter error.  The expression of $U^{t}(s)$ actually depends on how the Trotterization is performed, while our argument applies to general product formulas, thus we use the above definition as an instance. 

The total error of DAS, which is the fidelity distance between $A_{t}|\psi_{i}\rangle$ and $|\psi_{f}\rangle$, is divided into three parts. The first part comes from adiabatic evolution itself, it can be quantified by Lemma \ref{lemma:continu}. The other two stem from discretization and Trotterization steps. To simplify the question, we only study the case where the error caused by discretization is negligible comparing to the error from adiabatic theorem, which is to say:
\begin{gather}
\epsilon^{\prime}_{\text{adb}} \equiv \|P_{f} - A_{d}P_{i}A^{\dagger}_{d}\|\\
\epsilon_{\text{dis}} \equiv |\epsilon^{\prime}_{\text{adb}} - \epsilon_{\text{adb}}| \ll \epsilon_{\text{adb}}
\end{gather}
\begin{figure}

\includegraphics[scale=0.5]{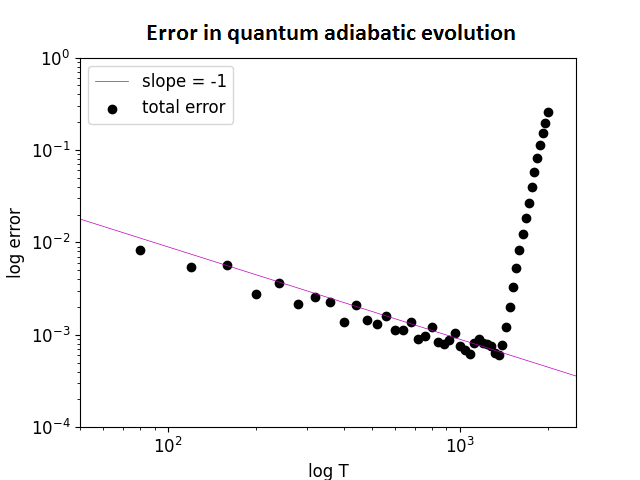}
\includegraphics[scale=0.5]{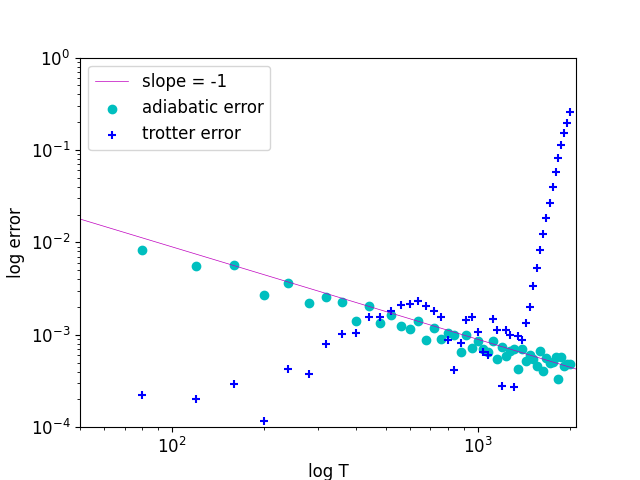}
\caption{Error scaling of DAS. The example is $H_{i} = -\sum_{i}X_{i}, H_{f} = -\sum_{i}(Z_{i} + Z_{i}Z_{i+1})$ on $N = 8$ sites without periodic boundary conditions. The initial state is the ground state of $H_{i}$.  The discretization number is fixed at $M = 2000$. The figures illustrate how $\epsilon^{\prime}_{\text{tot}}, \epsilon^{\prime}_{\text{adb}}$ and $\epsilon_{\text{tro}}$ scale with parameter $T$ ranging from $[M/50, M]$ evenly. Each point is a complete run of DAS and there are 50 of them in one line.  The vertical axis is shown on a $\log_{10}$ scale. The first figure is only about $\epsilon^{\prime}_{\text{tot}}$, the second figure is a combination of $\epsilon^{\prime}_{\text{adb}}$ and $\epsilon_{\text{tro}}$. As to $\epsilon^{\prime}_{\text{adb}}$, the overall scaling of $\mathcal{O}(T^{-1})$ indicates the correctness of prediction by adiabatic theorem.  Note that the Trotter error remains small when $T,M$ are comparable to each other. $\epsilon^{\prime}_{\text{tot}}$ is close to the summation of $\epsilon^{\prime}_{\text{adb}}$ and $\epsilon_{\text{tro}}$. The overall scaling is similar to that of $\epsilon^{\prime}_{\text{adb}}$, while after a turning point the error grows rapidly, and the curve matches with that of $\epsilon_{\text{tro}}$. The turning point is the place where the minimal error is reached. Approximately it's $0.75M$. }
\label{fig:err}
\end{figure}
We focus on the Trotter error in DAS, \begin{equation}
\epsilon_{\text{tro}} \equiv \|A_{d}P_{i}A^{\dagger}_{d} - A_{t}P_{i}A^{\dagger}_{t}\| \le \|A_{d} - A_{t}\|
\end{equation}
which has previously been bounded in terms of operator norm : $\|A_{d} - A_{t}\| = \mathcal{O}(T^{2}/M)$~\cite{barends2016digitized}. However, as indicated by the adiabatic theorem, all we need to control is the fidelity error, and by building on the techniques in Corollary \ref{coro:eigen} we will demonstrate that $\epsilon_{\text{tro}}$ has a tighter upper bound that differs from $\|A_{d} - A_{t}\|$. We associate a time-evolving effective Hamiltonian $\WH(s)$  with $U^{t}(s)$,
\begin{equation}
\WH(s) \equiv i\log(U^{t}(s))/\bar{\delta t}
\end{equation}
Notice that at the beginning point $s=0$ and end point $s=1$, $\WH(s)$ is the same as $H_{i}$ and $H_{f}$. So the adiabatic evolution under $\WH(s)$ can also transform $|\psi_{i}\rangle$ to $|\psi_{f}\rangle$. Naturally the total error should also be quantified with adiabatic theorem. 
\begin{gather}
\epsilon^{\prime}_{\text{tot}} \equiv \|P_{f} - A_{t}P_{i}A^{\dagger}_{t}\|
\end{gather}
Again, we require that the error caused by discretization is negligible:
\begin{gather*}
\epsilon_{\text{tot}} \equiv \|P_{f} - \widetilde{A}(T)P_{i}\widetilde{A}^{\dagger}(T)\|\\
\widetilde{A}(T) \equiv \exp_{\mathcal{T}}\left(-iT\int_{0}^{1}\WH(s)ds\right)\\
\widetilde{\epsilon}_{\text{dis}} \equiv |\epsilon^{\prime}_{\text{tot}} - \epsilon_{\text{tot}}| \ll \epsilon_{\text{tot}}
\end{gather*}
Here $\epsilon_{\text{tot}}$ is the adiabatic error of evolution under $\WH(s)$, which is also the continuous limit of $\epsilon^{\prime}_{\text{tot}}$. Using the adiabatic theorem we have:
\begin{equation}
\epsilon_{\text{tot}} \le \mathcal{G}(T,\WH)
\end{equation}

Combining with Lemma \ref{lemma:magnus}, Lemma \ref{lemma:continu} and the triangle inequality $\epsilon_{\text{tro}} \le \epsilon^{\prime}_{\text{adb}} + \epsilon^{\prime}_{\text{tot}}$, we prove the following proposition in Appendix F.
\begin{prop}
Consider a digital adiabatic evolution with initial Hamiltonian $H_{i}$, final Hamiltonian $H_{f}$, total evolution time $T$, and discretization number $M$. If the discretization error is negligible comparing to the continuous limit then
\begin{equation}
\epsilon_{dis} \ll \epsilon_{adb},\quad \widetilde{\epsilon}_{dis} \ll \epsilon_{tot}
\label{equ:c2}
\end{equation}
then the Trotter error has upper bound:
\begin{equation}
\epsilon_{\text{tro}} = \mathcal{O}(\mathcal{G}(T,H) + \mathcal{G}(T,\WH))
\end{equation}
with $\mathcal{G}$ introduced in Lemma \ref{lemma:continu}. Define:
\begin{gather*}
C_{0}\equiv\|H_{i}\|,\quad C_{1}\equiv\|[H_{i},H_{f}]\|\\
D\equiv\|H_{i}-H_{f}\|
\end{gather*}
when $D/\lambda \gg 1$, where $\lambda$ is the lower bound of spectral gap during adiabatic evolution under $H(s)$ and $\WH(s)$, and $(C_{0} + 3D/2)T<M/4$, we obtain:
\begin{equation}
\epsilon_{\text{tro}} = \mathcal{O}\left(\frac{D^{2}}{T\lambda^{3}}\right) + \mathcal{O}\left(\frac{C_{1}^{2}T}{M^{2}\lambda^{3}}\right)
\end{equation}
\label{prop:1}
\end{prop}
From Proposition \ref{prop:1} we already derive a result different from $\|A_{d} - A_{t}\|$. Numerical results (see Fig. \ref{fig:err}) seem to indicate that $\epsilon_{\text{tro}}$ is close to $\epsilon^{\prime}_{\text{tot}} - \epsilon^{\prime}_{\text{adb}}$, which means this result can be further improved. Also this result only works for $T/M = \mathcal{O}(N^{-1})$.

Though taken for granted in most works, the criterion for Eq : (\ref{equ:c2}) to hold is critical to the error analysis. This question has its independent interest. The discretized adiabatic evolution has been the subject of some analysis~\cite{ambainis2004elementary}, but we believe the result can be intrinsically improved and thus leave it for future work.

\section{Applications}

\subsection{Quantum Phase Estimation}
The QPE algorithm constructs a quantum circuit to detect the phase $\theta$ of a unitary operator: $U|\psi\rangle = e^{i\theta}|\psi\rangle$. 
For an exact QPE algorithm, the measurement outcome is the integer $a$ closest to $2^{l}\theta$, where $l$ is the size of the first register. Since it's unlikely for $2^{l}\theta$ to be an integer, there is an inherent error $\xi = \mathcal{O}(2^{-l})$ in this algorithm. The probability of measuring the value closest to the true $\theta$ is at least $4/\pi^{2}$~\cite{nielsen2002quantum}.

The influence of the Trotter error comes from two aspects.  Again we regard the Trotterized evolution operator as an exact evolution operator under the effective Hamiltonian $\widetilde{H}$. This effective Hamiltonian has an eigenstate $|\widetilde{\psi}\rangle$, which is very close to initial eigenstate $|\psi\rangle$ : $|\psi\rangle = \sqrt{1-p}|\widetilde{\psi}\rangle + \sqrt{p}|\widetilde{\psi}^{\perp}\rangle$. As a result, the final phase detected should be $\widetilde{\theta}$ and the success rate should be decreased by a factor of $(1-p)$. However, since usually  $p$ is much smaller comparing to 1, this change in success rate is almost negligible. 

More importantly, the Trotter error in phase $\delta\theta = |\widetilde{\theta} - \theta|$ should satisfy $\delta\theta < \xi$, otherwise the phase error caused by Trotterization will be detected. This relation gives us a constraint on the Trotter error:
\begin{equation}
|\widetilde{E} - E|t \le \xi
\end{equation}
In QPE, $\theta$ should be set to be close to 1 to avoid wasting the accuracy provided by the first register, thus $t = \mathcal{O}(1/
|E|)$. However, we can only guess about $E$ before the algorithm. Here we use $t_{0}$ to denote an appropriate choice of time scale in $U$. Thus
\begin{equation}
L = \mathcal{O}\left(\frac{t_{0}}{\delta t}\right)
\end{equation}
where $L$ is the Trotterization number. 

The following result follows from Lemma \ref{lemma:spe}.
\begin{corollary}
Suppose there's a quantum circuit performing QPE, the size of the first register is $l$ thus the inherent error is $\xi = \mathcal{O}(2^{-l})$. The unitary operator is $U = e^{-iHt_{0}}$ where $H$satisfies one of the conditions from Lemma \ref{lemma:spe} in the previous section. To guarantee that the Trotter error in phase is smaller than the inherent error $\xi$, we have:
\begin{gather}
\delta t = \mathcal{O}\left(\frac{1}{N}\sqrt{\frac{\xi}{t_{0}}}\min\{1,\lambda\}\right)\\
L = \mathcal{O}\left(N\sqrt{\frac{t_{0}^{3}}{\xi}}\max\left\{1,\frac{1}{\lambda}\right\}\right)\\
\text{Circuit Depth} = \mathcal{O}\left(\sqrt{\frac{N^{3}t_{0}^{3}}{\xi^{3}}}\max\left\{1,\frac{1}{\lambda}\right\}\right)
\end{gather}
$\lambda$ is the lower bound of spectral gap between initial state $|\psi\rangle$ and its neighboring eigenstates.
\end{corollary}
As a comparison, in general case with $|\widetilde{E} - E| = \mathcal{O}(N\delta t)$, the final circuit depth is $\mathcal{O}(N^{2}t^{2}_{0}/\xi^{2})$.

\subsection{Optimizing Digital Adiabatic Simulation}


Although a better upper bound of $\epsilon_{\text{tro}}$ has been derived in Proposition $\ref{prop:1}$, it is really $\epsilon^{\prime}_{\text{tot}}$ rather than $\epsilon_{\text{tro}}$ that is the most relevant quantity, for it doesn't matter whether the error originates from numerical procedure or the finite size of $T$.  Motivated by this, here we elaborate on a different question: given a DAS task with $H_{i}$, $H_{f}$ and $|\psi_{i}\rangle$ specified, and a quantum computer with fixed circuit depth $M$, find the optimal $T$ to minimize the estimated total error $\epsilon^{\prime}_{\text{tot}}$. 

There exists a trade-off between Trotter error and adiabatic error: on one hand, $T$ can't be too small as the adiabatic error is inversely proportional to $T$; on the other hand, the error caused by Trotterization increases with $T$ (for a fixed depth $M$). The trade-off is balanced when the two errors are of the same magnitude. We use $T_{c}$ to denote the balanced point, which is exactly the turning point in Fig. \ref{fig:err}, if our estimation of $\epsilon_{\text{tot}}$ is accurate enough. The value of $T_{c}$ depends on the estimation of $\epsilon^{\prime}_{\text{tot}}$. In Proposition \ref{prop:1} we have derived an upper bound for $\epsilon^{\prime}_{\text{tot}}$ and denote it with the function $\mathcal{G}(T,\WH)$.  The critical value of $T_c$ and the optimal error are defined by
\begin{equation}
\epsilon_{\text{opt}} \equiv \inf_{T}\mathcal{G}(T,\WH),\quad T_{c}\equiv \arg\min \mathcal{G}(T,\WH).
\end{equation}
The next Corollary indicates that $T_{c}$ is proportional to $M$ and their optimal ratio determines the gate complexity of DAS. The result follows from Proposition \ref{prop:1}.
\begin{corollary}[Optimal choice of $T$]
In the setting of Proposition \ref{prop:1}, we intend to perform a digital adiabatic evolution on a quantum device with circuit depth $M$. If $(8C_{0} + 12D)D \le 3C_{1}$, then the upper bound of $\epsilon^{\prime}_{\text{tot}}$ reaches its minimal value at:
\begin{equation}
T_{c} = \frac{2MD}{3C_{1}}
\end{equation}
Accordingly,
\begin{equation}
\epsilon_{\text{opt}} = \mathcal{O}\left(\frac{DC_{1}}{M\lambda^{3}}\right)
\end{equation}
Alternatively, to achieve an error $\epsilon$ we can keep the ratio $T_{c}/M$ fixed and take the depth of quantum circuit $M$ to be:\begin{equation}
\text{Circuit Depth} = \mathcal{O}\left(\frac{NDC_{1}}{\lambda^{3}\epsilon}\right)
\end{equation}
$N$ is the width of quantum circuit and $\lambda$ is the lower bound of spectral gap.
\label{coro:aqc}
\end{corollary}
However, the $T_{c}$ derived still deviates a lot from the turning point in Fig. \ref{fig:err}, because our estimation of $\epsilon^{\prime}_{\text{tot}}$ is much larger than the actual value. A tighter bound for $\epsilon^{\prime}_{\text{tot}}$ would lead to a more accurate estimate of $T_{c}$.   Additional factors may contribute to the location of the turning point; for example, if the spectral gap of $\widetilde{H}(s)$ closes as $\delta t$ gets larger, the application of the adiabatic theorem will eventually fail.

\section{Conclusion $\&$ Outlook}

Our main contribution is the observation that refined estimation of Trotter error can be established from effective Hamiltonian $\WH$. When the initial state is an eigenstate, we first relate the fidelity error and phase error to the spectrum analysis of $\WH$, and find that during evolution, most error accumulates in the phase. Further, if the leading perturbation term of $\WH$ vanishes in the eigenbasis of $H$, the Trotter error in energy is reduced from $\mathcal{O}(\delta t)$ to $\mathcal{O}(\delta t^{2})$, which results in improvement of QPE.  We remark that the improvement in Trotter step size with phase estimation error $\epsilon$ from $\epsilon$ to $\epsilon^{1/2}$ can be compared to the step size of $\epsilon^{1/2}$ that is obtained from the use of a  second-order product formula approximation.  Similar results apply to other QPE methods such as robust phase estimation~\cite{russo2020evaluating} (see Appendix G), as long as the Trotterized unitary operator is used. Finally, we show that the spectral analysis method is particularly suitable to analyzing Trotter error in DAS, and we demonstrate how consideration of the various types of error leads to an optimal time parameter $T_{c}$ for DAS when the circuit depth of the quantum circuit is fixed.

There are many targets to pursuit in future work.  Here we have only studied Trotter error for the 1st order product formula, and it will be interesting to see whether similar properties exist for higher-order product formulas.   One may also seek examples of $f \ll \Delta^{2}$ in time-dependent Hamiltonian situation, in which the initial state is the eigenstate of the initial Hamiltonian, generalizing our results for DAS. Third, numerically the Trotter error in DAS is more close to $\epsilon^{\prime}_{\text{tot}} - \epsilon^{\prime}_{\text{adb}}$ rather than the summation of them, we believe with detail analysis this improvement can be achieved. The forth point is the criterion for the time step scale $T/M$ in DAS that keeps the error caused by discretization negligible. Finally, we would like to try the effective Hamiltonian idea in other quantum simulation algorithms~\cite{campbell2019random,low2019hamiltonian,low2017optimal,berry2015simulating}.

\section{Acknowledgement}

We thank Rolando Somma and Burak \c{S}ahino\u{g}lu for helpful discussions.   This material is based upon work supported by the U.S. Department of Energy, Office of Science, National Quantum Information Science Research Centers, Quantum Systems Accelerator (QSA).

\appendix
\renewcommand{\appendixname}{Appendix~\Alph{section}}

\bibliographystyle{unsrt}
\bibliography{references}

\pagebreak
\widetext

\subsection{Appendix A : Proof of Eq : (\ref{equ:fandtheta})} 
\label{app:A}
First we will quantify the region for $f$ and $\theta$ in Eq : (\ref{equ:setup}). We know that $f\in[0,1]$ by definition. As to $\theta$, consider the inner product:
\begin{equation*}
\langle\psi|U^{\dagger}(t)T(\delta t)^{L}|\psi\rangle = 1 + \langle\psi|U^{\dagger}(t)\hat{\Delta}|\psi\rangle
\end{equation*}
Since $\Delta \equiv \|T(\delta t)^{L} - U(t)\| \le 1/\sqrt{2}$:
\begin{equation*}
|\sin\theta| \le |\langle\psi|U^{\dagger}(t)\hat{\Delta}|\psi\rangle| \le \Delta \le 1/\sqrt{2}. 
\end{equation*}
Thus $|\theta| \in [-\pi/4,\pi/4]$. In this region, $\theta$ satisfies:
\begin{equation*}
1 - \frac{\theta^{2}}{2} \le \cos\theta \le 1 - \frac{\theta^{2}}{4}
\end{equation*}
Similarly,
\begin{equation*}
1 - f \le \sqrt{1-f} \le 1 - \frac{f}{2}
\end{equation*}
The Euclidean distance between two evolved states $\mathcal{E} = \|T(\delta t)^{L}|\psi\rangle - U(t)|\psi\rangle\|_{2}$ can be exactly represented with $f$ and $\theta$.
\begin{equation}
\mathcal{E}^{2} = 2 - 2\sqrt{1-f}\cos\theta
\end{equation}
From above estimation, we can quantify the upper/lower bound of $\mathcal{E}$
in terms of $f$ and $\theta$:
\begin{gather*}
2 - 2\sqrt{1-f}\cos\theta \ge 2 - 2(1-\frac{f}{2})(1-\frac{\theta^{2}}{4}) \ge f + \frac{\theta^{2}}{4}\\
2 - 2\sqrt{1-f}\cos\theta \le 2 - 2(1-f)(1-\frac{\theta^{2}}{2}) \le 2f + \theta^{2}
\end{gather*}
Combine them together:
\begin{equation*}
f + \frac{\theta^{2}}{4} \le \mathcal{E}^{2} \le 2f + \theta^{2}
\end{equation*}

\section{Appendix B : Proof of Lemma \ref{lemma:magnus}}

Here we exhibit how to compare the spectrum and eigenstates of effective Hamiltonian $\{\widetilde{E}_{k}, |\wpk\rangle\}$ with that of original Hamiltonian $\{E_{k},|\pk\rangle\}$ rigorously. The method has already been established in~\cite{tran2020faster}. Magnus expansion sets up connection between two types of exponential operators:
\begin{gather*}
\exp_{\mathcal{T}}\left[-i\int_{0}^{t}E(\tau)d\tau\right] = \exp(-iX), \quad
-iX = \sum_{j}\Omega_{j}\\
\Omega_{j} = \frac{1}{j^{2}}\sum_{\sigma\in S_{j}}\frac{(-1)^{d}(-i)^{j}}{C^{d}_{j-1}}\cdot
\int_{0}^{t}dt_{1}\cdots\int_{0}^{t_{n-1}}dt_{n}[E(t_{1}),\cdots,[E(t_{n-1}),E(t_{n})],\cdots]
\end{gather*}
where $S_{j}$ is permutation group and $d$ is a constant related to a permutation. This infinite series $\{\Omega_{j}\}$ is called Magnus expansion. It has been proved that:
\begin{equation*}
\|\widetilde{H} - H\| \le \frac{\alpha \delta t}{2} + \delta t^{2}(\beta + 32\alpha \|H\|)
\end{equation*}
With parameters defined in Lemma \ref{lemma:magnus}. There are lots of constrains for $\delta t$ in the derivation, we summarize these constraints by setting $\delta t = \mathcal{O}(N^{-1})$. As to $\|\widetilde{H}^{\prime}\|$, previous results show:
\begin{gather*}
\widetilde{H} = H - \frac{i}{2}\hat{\alpha}\delta t + \frac{1}{\delta t}\int_{0}^{\delta t}dx \hat{\beta}(x) + \frac{1}{\delta t}\sum_{n=2}\Omega_{n}(\delta t)\\
\hat{\alpha} = \sum_{l>m}[H_{l},H_{m}],\quad \|\hat{\beta}(x)\| \le x^{2}\beta\\
\|\Omega_{n}\| \le \sum_{\sigma}\int\cdots\int dt_{1}\cdots dt_{n} \mathcal{C}_{n}\\
\mathcal{C}_{n} \le 2^{n}(\|H\| + \alpha \delta t + \beta \delta t^{2})^{n-1}(\alpha \delta t + \beta \delta t^{2})
\end{gather*}
When we upper bound the derivatives, we separate the above expression into three parts and consider the worst case: (we write $\delta t$ as $t$ temporarily for abbreviation)
\begin{equation*}
\|\left(\frac{1}{t}\int_{0}^{t}\hat{\beta}(x)dx\right)^{\prime}\| \le \frac{1}{t}\|\hat{\beta}(t)\| + \frac{1}{t^{2}}\int_{0}^{t}\|\hat{\beta}(x)\|dx \le \frac{4}{3}\beta t
\end{equation*}
Next we will prove $\|\sum_{n=2}\Omega^{\prime}_{n}(t)\| \le \omega t^{2}$ to directly use the above scenario.
\begin{equation*}
\begin{aligned}
\|\sum_{n=2}\Omega^{\prime}(t)\| &\le \sum_{n=2}n!\frac{t^{n-1}}{(n-1)!}\mathcal{C}_{n}\\
&\le \sum_{n=2}nt^{n-1}2^{n}(\|H\| + \alpha t + \beta t^{2})^{n-1}(\alpha t + \beta t^{2})\\
&= 2t(\alpha + \beta t)\sum_{n=2}n(2\|H\|t + 2\alpha t^{2} + 2\beta t^{3})^{n-1}\\
&= 2t(\alpha + \beta t)g(2\|H\|t + 2\alpha t^{2} + 2\beta t^{3})
\end{aligned}
\end{equation*}
where $g(y) = (y^{2}/(1-y))^{\prime} < 4y + 4y^{2} < 8y$ when $y<1/2$. Thus:
\begin{equation*}
\omega = 128\alpha\|H\|
\end{equation*}
Replace $t$ with $\delta t$. Finally:
\begin{equation*}
\|\widetilde{H}^{\prime}(\delta t)\| \le \frac{\alpha}{2} + \frac{4}{3}(\beta + 128\alpha\|H\|)\delta t
\end{equation*}

\subsection{Appendix C : Proof of Cororllary \ref{coro:eigen}}

This result is a special case of Theorem \ref{coro:multi}. Here we prove it in a more direct way. Calculate the inner product between $U(t)|\pk\rangle$ and $T(\delta t)^{L}|\pk\rangle$ using spectral decomposition $\WH = \sum_{j}\widetilde{E}_{j}\widetilde{P}_{j}$:
\begin{equation*}
\begin{aligned}
\sqrt{1-f} e^{i\theta} &= \langle\pk|U^{\dagger}(t)T(\delta t)^{L}|\pk\rangle\\ 
&= \sum_{j}e^{i(E_{k} - \widetilde{E}_{j})t}|\langle\pk|\widetilde{\psi}_{j}\rangle|^{2}\\
&= e^{i(E_{k} - \widetilde{E}_{k})t} |\langle\pk|\widetilde{\psi}_{k}\rangle|^{2} + \sum_{j\neq k} e^{i(E_{k} - \widetilde{E}_{j})t}|\langle \pk|\widetilde{\psi}_{j}\rangle|^{2}\\
&= e^{i(E_{k} - \widetilde{E}_{k})t}\left[|\langle\pk|\wpk\rangle|^{2} + \sum_{j\neq k}e^{i(\widetilde{E}_{k} - \widetilde{E}_{j})t}|\langle\pk|\widetilde{\psi}_{j}\rangle|^{2}\right]
\end{aligned}
\end{equation*}
Define $\varepsilon$ and $\eta$ as:
\begin{gather*}
\varepsilon = 1 - |\langle\pk|\wpk\rangle|^{2},\quad
\eta = \sum_{j\neq k}e^{i(\widetilde{E}_{k} - \widetilde{E}_{j})t}|\langle\pk|\widetilde{\psi}_{j}\rangle|^{2}
\end{gather*}
Then:
\begin{gather*}
1 - f = |1 - \varepsilon + \eta|^{2},\quad
\theta = (E_{k} - \widetilde{E}_{k})t + \text{Arg}(1 - \varepsilon + \eta)
\end{gather*}
The complex number $\eta$ is upper bounded by $|\eta| \le \varepsilon$ and the equality is satisfied when all the phases $(\widetilde{E}_{j} - \widetilde{E}_{k})t$ differ by $2\pi j$, hence:
\begin{gather*}
f = \mathcal{O}(1-|\langle \pk|\wpk\rangle|^{2}), \quad\theta = \mathcal{O}(|\widetilde{E}_{k} - E_{k}|t)
\end{gather*}
Combine with Lemma \ref{lemma:magnus}:
\begin{equation*}
f = \mathcal{O}\left(\frac{h^{2}\delta t^{2}}{\lambda^{2}}\right),\quad |\theta| = \mathcal{O}(Lh\delta t^{2})
\end{equation*}
Of course, $f = \theta = 0$ when $L=0$. However, in this upper bound $f$ is irrelevant to $L$. To fix this, notice we have another upper bound for $f$ from Eq : (\ref{equ:fandtheta}):
\begin{gather*}
f \le \mathcal{E}^{2} \le \Delta^{2} = \mathcal{O}(L^{2}h^{2}\delta t^{4})\\
\end{gather*}

\section{Appendix D : Proof of Eq : (\ref{equ:norm_to_fid}) and Theorem \ref{coro:multi}}

Consider the norm distance between two projectors $\delta p = \|P_{x} - P_{y}\|, P_{x} = |x\rangle\langle x|, P_{y} = |y\rangle\langle y|$, one observation is that this matrix is normal, hence its norm is the largest value of absolute eigenvalues of the following matrix:
\begin{gather*}
|y\rangle = a|x\rangle + b|x^{\perp}\rangle,\quad
\Delta P = \begin{bmatrix}
1 - a^{2} & -ab\\
-ab & -b^{2}
\end{bmatrix}
\end{gather*}
Although $\Delta P$ is only the representation of projector difference in a subspace spanned by $\{|x\rangle,|y\rangle\}$, other dimensions won't effect the spectrum of it. The parameters $a,b$ can be set as non-negative real numbers for there are two degrees of freedom on the phases of $|y\rangle$ and $|x^{\perp}\rangle$. Also $a^{2}$ is the fidelity between two states. 

Solve this matrix, the largest absolute value of $\Delta P$ is:
\begin{gather*}
\delta p =  b = \sqrt{1 - |\langle x|y\rangle|^{2}}
\end{gather*} 

We also also extend this result to multi-state projectors. Suppose we have two projectors $P_{A}$ and $P_{B}$ that each corresponds to an $m$-dimensional subspace. $A$,$B$ are the invariant spaces of the two projectors. To make the problem meaningful, they should be close to each other in the sense that $\|P_{A} - P_{B}\| < 1$. In another word, $A$ and $B$ share $m-1$ dimensions. Then we can re-choose two sets of basis in $A$ and $B$ such that only one element is different. Then the problem is reduced to the previous one. Based on this observation:
\begin{equation*}
\min_{|\phi_{a}\rangle\in \mathcal{H}_{A}}\text{tr}(P_{B}|\phi_{a}\rangle\langle\phi_{a}|) = 1 - \|P_{A} - P_{B}\|^{2}
\end{equation*} 
Similar result can be extended to mixed state, since the extreme status corresponds to pure state:
\begin{equation}
\min_{\rho_{a}\in \mathcal{H}_{A}}\text{tr}(P_{B}\rho_{a}) = 1 - \|P_{A} - P_{B}\|^{2}
\label{equ:lb_mix}
\end{equation}
In Theorem \ref{coro:multi}, we have an original projector $P$ that is composed of $m$ eigenstate projectors, and an ``effective" projector $\widetilde{P}$ which is very close to $P$. Their norm distance has been well-quantified. Write:
\begin{equation*}
\delta p \equiv \|\widetilde{P} - P\|
\end{equation*}
Our initial state $\rho$ is a mixed state inside the space of $P$, which means $\text{tr}(\rho P) = 1$. In Eq : (\ref{equ:lb_mix}) we have proved that:
\begin{equation*}
\text{tr}(\rho\widetilde{P}) \ge 1 - \delta p^{2}
\end{equation*}
Now we want to quantify the leakage rate:
\begin{equation*}
1 - \text{tr}(\widetilde{U}\rho\widetilde{U}^{\dagger} P) = 1 - \text{tr}(\rho\widetilde{U}^{\dagger} P\widetilde{U})
\end{equation*}
where $\widetilde{U}$ is the Trotterized evolution operator. Easy to see that $[\widetilde{U},\widetilde{P}] = 0$. Therefore:
\begin{gather*}
\|\widetilde{P} - \widetilde{U}^{\dagger}P\widetilde{U}\| = \|\widetilde{P} - P\| = \delta p\\
\widetilde{U}^{\dagger}P\widetilde{U} = P + (\widetilde{P} - P) + (\widetilde{U}^{\dagger}P\widetilde{U} - \widetilde{P})\\
\|\widetilde{U}^{\dagger}P\widetilde{U} - P\| \le 2\delta p
\end{gather*}
Use the same argument in Eq : (\ref{equ:lb_mix}) we derive:
\begin{gather*}
\text{tr}(\rho\widetilde{U}^{\dagger} P\widetilde{U}) \ge 1 - 4\delta p^{2}\\
1 - \text{tr}(\rho\widetilde{U}^{\dagger} P\widetilde{U}) = \mathcal{O}(\delta p^{2})
\end{gather*}
The leakage rate has order $\mathcal{O}(\delta p^{2})$.

\section{Appendix E : Proof of Lemma \ref{lemma:spe}}

Define $H(s)$ as:
\begin{equation*}
H(s) \equiv H + sV,\quad V\equiv\frac{i}{2}\sum_{l>m}[H_{l},H_{m}]
\end{equation*}
with eigenstates and eigenvalues:
\begin{equation*}
H(s)|\psi(s)\rangle = E(s)|\psi(s)\rangle
\end{equation*}
The actual effective Hamiltonian satisfies $\WH = H(\delta t) + V_{2}$. Use the method in Lemma \ref{lemma:magnus} we can prove $\|V_{2}\| = \mathcal{O}(\delta t^{2}(\beta + 32\alpha\|H\|))$. We focus on the error in energy caused by $H(s)$ first. Lemma \ref{lemma:spe} states that $\langle\psi(0)|V|\psi(0)\rangle = 0$. To exploit this condition, define fidelity distance $f(s)$ from the following equation:
\begin{equation*}
|\psi(s)\rangle = \sqrt{1-f(s)}|\psi(0)\rangle + \sqrt{f(s)}|\psi(0)^{\perp}\rangle
\end{equation*}
We don't need $e^{i\theta(s)}$ as there's a degree of freedom on the phase of $|\psi(s)\rangle$. Use the method in Corollary \ref{coro:eigen} we have:
\begin{gather*}
1 - |\langle\psi(0)|\psi(s)\rangle|^{2} = f(s) = \|P(s) - P(0)\|^{2}\\
\|P(s) - P(0)\| \le s\max_{x\in[0,s]} \|P^{\prime}(x)\|,\max_{x\in[0,s]}\|P^{\prime}(x)\| \le \frac{\|V\|}{\lambda} 
\end{gather*}
Hence, $f(s) \le \frac{\|V\|^{2}s^{2}}{\lambda ^{2}}$. Since $E(s) = \langle\psi(s)|H(s)|\psi(s)\rangle$ and $\langle\psi(0)|V|\psi(0)\rangle = 0$:
\begin{equation*}
\begin{aligned}
E(s) - E(0) =& \langle\psi(s)|H(s)|\psi(s)\rangle - \langle\psi(0)|H|\psi(0)\rangle\\
 =& f(s)(\langle\psi(0)^{\perp}|H|\psi(0)^{\perp}\rangle - \langle\psi(0)|H|\psi(0)\rangle)+f(s)s\langle\psi(0)^{\perp}|V|\psi(0)^{\perp}\rangle\\
& + s\sqrt{(1-f(s))f(s)}(\langle\psi(0)|V|\psi(0)^{\perp}\rangle + c.c)\\
\end{aligned}
\end{equation*}
Here we use the following argument for upper bound:
\begin{gather*}
|\phi\rangle = \frac{1}{\sqrt{2}}(|\phi\rangle + |\psi^{\perp}\rangle)\\
\langle\phi|V|\phi\rangle = \frac{1}{2}(\langle\psi|V|\psi\rangle + \langle\psi^{\perp}|V|\psi^{\perp}\rangle
+ \langle\psi|V|\psi^{\perp}\rangle + c.c)\\
\langle\psi|V|\psi^{\perp}\rangle + c.c \le 2\|V\|
\end{gather*}
Finally:
\begin{equation*}
|E(s) - E(0)| \le \frac{\|V\|^{2}\cdot\|H\|s^{2}}{\lambda^{2}} + \frac{2\|V\|^{2}s^{2}}{\lambda} + \frac{\|V\|^{3}s^{3}}{\lambda^{2}}
\end{equation*}
When $\|V\|s < \lambda$ and $\|H\| > 2\lambda$, only the first term remains. Thus:
\begin{equation}
|E(\delta t) - E(0)| = \mathcal{O}\left(\frac{\|V\|^{2}\cdot\|H\|\delta t^{2}}{\lambda^{2}}\right)
\end{equation}
Quantify the error caused by $V_{2}$ with Weyl's inequality:
\begin{equation*}
\begin{aligned}
|\widetilde{E} - E(0)| & \le |\widetilde{E} - E(\delta t)| + |E(\delta t) - E(0)|\\
& = \mathcal{O}(\delta t^{2}(\beta + 32\alpha\|H\|)) + \mathcal{O}\left(\frac{\|V\|^{2}\cdot\|H\|\delta t^{2}}{\lambda^{2}}\right)\\
& = \mathcal{O}\left(N^{2}\delta t^{2}\max\left\{1,\frac{1}{\lambda^{2}}\right\}\right)
\end{aligned}
\end{equation*}

\section{Appendix F : Proof of Proposition \ref{prop:1} and Corollary \ref{coro:aqc}}

The content in this Appendix is nothing more than combining Lemma \ref{lemma:magnus} with Lemma \ref{lemma:continu}. To quantify $\mathcal{G}(T,\WH)$, our target is the derivative of $\WH(s)$ defined as:
\begin{equation*}
\WH(s) \equiv i\log\left(\exp(-it(1-s)H_{i})\exp(-itsH_{f})\right)/t
\end{equation*}
We use $t$ to denote $\delta t$ for simplicity in this Appendix. $\WH$ satisfies:
\begin{equation*}
e^{-it\WH(s)} = e^{-itH_{i}}\cdot\left(e^{istH_{i}}e^{-istH_{f}}\right) = e^{-itH_{i}}e^{-itG(s)}
\end{equation*}
We separate the calculation of $\WH$ into two parts. In the first part we take derivative with respect to $t$, and represent $\WH$ with $H_{i}$ and $G(s)$; next we calculate $G(s)$ by taking derivatives with respect to $s$ instead. The procedure is similar to that of Appendix B. The following special functions are used in the estimation:
\begin{gather*}
\mathcal{F}_{0}(x) = \sum_{j=0}^{\infty}x^{j} = \frac{1}{1 - x}\\
\mathcal{F}_{1}(x) = \sum_{j=1}^{\infty}\frac{1}{j}x^{j-1} = - \ln(1-x)/x\\
\mathcal{F}_{2}(x) = \sum_{j=2}^{\infty}\frac{1}{j^{2}}x^{j} = \int_{0}^{x}-\ln(1-x^{\prime})dx^{\prime}
\end{gather*}
They have the following upper bounds when $0 \le x\le 1/2$:
\begin{gather*}
\mathcal{F}_{0} \le 1 + 2x,\quad \mathcal{F}_{1} \le 1 + x,\quad \mathcal{F}_{2} \le \frac{x^{2}}{2}(1+x)
\end{gather*}
We express $\WH$ in terms of $G$ first. Under the setting of Magnus expansion:
\begin{gather*}
-it\WH(s) = \sum_{j=1}^{\infty}\Omega_{j}(s,t),\quad E(s,t) = -ie^{-itH_{i}}(H_{i}+G(s,t))e^{itH_{i}}\\
\Omega_{j}(s,t) = \frac{1}{j^{2}}\sum_{\sigma\in S_{j}}\frac{(-1)^{d}}{C^{d}_{j-1}}\int_{0}^{t}dt_{1}\cdots\int_{0}^{t_{j-1}}dt_{j}[E(t_{1}),\cdots[E(t_{j-1}),E(t_{j})]]
\end{gather*}
$d$ is a constant relevant to permutation $\sigma$. As to $\WH^{\prime}(s)$, the integrand in $\Omega^{\prime}_{j}(s)$ becomes summation over $j$ different commutators; while for $\WH^{\prime\prime}(s)$, the integrand of $\Omega_{j}^{\prime\prime}(s)$ contains $j(j-1)$ pairs of $E^{\prime}$ or $j$ $E^{\prime\prime}$. Thus:
\begin{gather*}
\|\WH^{\prime}(s)\| \le \|G^{\prime}\|\cdot\mathcal{F}_{1}\left(2(\|H_{i} + G\|)t\right)\\
\|\WH^{\prime\prime}(s)\| \le \|G^{\prime\prime}\|\cdot\mathcal{F}_{1}\left(2(\|H_{i} + G\|)t\right) + 2\|G^{\prime}\|^{2}t\cdot\mathcal{F}_{0}(2\|H + G\|t)
\end{gather*}
The next step is to represent $G(s,t)$ with $H_{i}$ and $H_{f}$, and we consider $s$ as variable instead.
\begin{gather*}
\frac{d}{ds}e^{-itG(s,t)} = \widetilde{E}(s,t)e^{-itG(s,t)},\quad \widetilde{E}(s,t) = e^{istH_{i}}(itH_{i} - itH_{f})e^{-istH_{i}}\\
G(s,t) = its(H_{i} - H_{f}) + t^{2}\int_{0}^{s}ds^{\prime}\int_{0}^{s^{\prime}}ds^{\prime\prime}e^{its^{\prime\prime}H_{i}}[H_{i},H_{f}]e^{-its^{\prime\prime}H_{i}} + \sum_{j=2}^{\infty}\widetilde{\Omega}_{j}(s,t)\\
\widetilde{\Omega}_{j}(s,t) = \frac{1}{j^{2}}\sum_{\sigma\in S_{j}}\frac{(-1)^{d}}{C^{d}_{j-1}}\int_{0}^{s}ds_{1}\cdots\int_{0}^{s_{j-1}}ds_{j}[E(s_{1}),\cdots,[E(s_{j-1}),E(s_{j})]]
\end{gather*}
To abbreviate the result, define:
\begin{equation*}
C_{0} \equiv \|H_{i}\|,\quad C_{1} \equiv \|[H_{i},H_{f}]\|,\quad C_{2} \equiv \|[H_{i},[H_{i},H_{f}]]\|,\quad D \equiv \|H_{i} - H_{f}\|
\end{equation*}
Notice that we treat $\widetilde{\Omega}_{1}(s,t)$ separately to derive a better result. The norm of $G(s,t), G^{\prime}(s,t)$ and $G^{\prime\prime}(s,t)$ can be bounded by: 
\begin{gather*}
\|G(s,t)\| \le sD + \frac{1}{2t}\mathcal{F}_{2}(2stD)\\
\|\frac{d}{ds}G(s,t)\| \le D + stC_{1}\mathcal{F}_{1}(2stD)\\
\|\frac{d^{2}}{ds^{2}}G(s,t)\| \le C_{1}t + 2C_{1}^{2}s^{2}t^{3}\mathcal{F}_{0}(2stD) + C_{2}st^{2}(\mathcal{F}_{1}(2stD) - 1)
\end{gather*}
Combine everything together:
\begin{gather*}
\|\WH^{\prime}(s)\| \le [D + C_{1}t(1+2Dt)][1 + (2C_{0}+3D)t]\\
\|\WH^{\prime\prime}(s)\| \le \left(C_{1}t + 4C_{1}^{2}t^{3} + 2DC_{2}t^{3}\right)[1 + (2C_{0} + 3D)t] + 2t\left(D +C_{1}t(1+2Dt) \right)^{2}\frac{1}{1 - (2C_{0} + 3D)t}
\end{gather*}
Here it's required that $tD < 1/4$ and $(\|H_{i}\| + \max\|G(s)\|)t < 1/4$. 

Put the above results in $\mathcal{G}(T,\WH)$:
\begin{equation*}
\begin{aligned}
\mathcal{G}(T,\WH) &= \frac{1}{T}\left(\frac{\|\WH^{\prime}(0)\|}{\widetilde{\lambda}(0)^{2}} + \frac{\|\WH^{\prime}(1)\|}{\widetilde{\lambda}(1)^{2}}\right) + \frac{1}{T}\int_{0}^{1}\frac{\|\WH^{\prime\prime}(s)\|}{\widetilde{\lambda}^{2}(s)} + 7\frac{\|\WH^{\prime}(s)\|^{2}}{\widetilde{\lambda}^{3}(s)}ds\\
&\le \frac{2}{T\lambda^{2}}\max\|\WH^{\prime}(s)\| + \frac{1}{T\lambda^{2}}\max\|\WH^{\prime\prime}(s)\| + \frac{7}{T\lambda^{3}}\max\|\WH^{\prime}(s)\|^{2}
\end{aligned}
\end{equation*}
$\lambda \equiv \inf_{s}\{\widetilde{\lambda}(s)\}$. 
In this upper bound, $\WH^{\prime\prime}(s)$ won't appear in the leading term. Given $N$ as the size of the system, roughly, the magnitudes of quantities are $t = \mathcal{O}(N^{-1}),D = \mathcal{O}(N),C_{k} = \mathcal{O}(N^{k+1})$. As a result, $\|\WH^{\prime}\| = \mathcal{O}(N), \|\WH^{\prime\prime}\| = \mathcal{O}(N)$, and the leading term of the upper bound is:
\begin{equation*}
\begin{aligned}
\epsilon_{\text{tot}} &\le \mathcal{O}\left(\frac{1}{T\lambda^{3}}\left(D + C_{1}t(1 + 2Dt)\right)^{2}[1 + (2C_{0} + 3D)t]^{2}\right)\\
&= \mathcal{O}\left(\frac{1}{T\lambda^{3}}\left(D + \frac{3C_{1}T}{2M}\right)^{2}\right)
\end{aligned}
\end{equation*}
The maximal of the above upper bound is reached at:
\begin{equation}
\frac{T_{c}}{M} = \frac{2D}{3C_{1}}
\end{equation}
We further requires that $(8C_{0} + 12D)D \le 3C_{1}$ to let $T_{c}$ satisfies $(\|H_{i}\| + \max\|G(s)\|)T < M/4$.

\section{Appendix G : Trotter Error in Robust Phase Estimation}

The idea of Robust Phase Estimation is, begin with two quantum states:
\begin{gather*}
|\alpha\rangle = \frac{1}{\sqrt{2}}\left(|0\rangle + |1\rangle\right)\\
|\beta\rangle = \frac{1}{\sqrt{2}}\left(|0\rangle + i|1\rangle\right)
\end{gather*}
The target is the phase difference $(E_{0} - E_{1})t$. It can be derived from the outcome of two measurements:
\begin{gather*}
P_{\alpha} = |\langle\alpha|U(t)|\alpha\rangle|^{2} = \frac{1}{2}(1 + \cos((E_{1} - E_{0})t))\\
P_{\beta} = |\langle\alpha|U(t)|\beta\rangle|^{2} = \frac{1}{2}(1 + \sin((E_{1} - E_{0})t))
\end{gather*}
Thus:
\begin{equation*}
\tan((E_{1} - E_{0})t) = \frac{2P_{\beta} - 1}{2P_{\alpha} - 1}
\end{equation*}
Now let's consider the Trotterized version:
\begin{gather*}
|0\rangle = \sqrt{1 - f_{0}}|\wz\rangle + \sqrt{f_{0}}|\wz^{\perp}\rangle\\
|1\rangle = \sqrt{1 - f_{1}}|\wo\rangle + \sqrt{f_{1}}|\wo^{\perp}\rangle\\
\wu |\wz\rangle = e^{-i\widetilde{E}_{0}t}|\wz\rangle\quad,\quad \wu |\wo\rangle = e^{-i\widetilde{E}_{1}t}|\wo\rangle
\end{gather*}
From here we have:
\begin{gather*}
\langle \alpha|\widetilde{U}(t)|\alpha\rangle  = \frac{1}{2}[ e^{-i\widetilde{E}_{0}t} + e^{-i\widetilde{E}_{1}t} + 2\sqrt{f_{1}}\text{Re}(\langle\wz|\wo^{\perp}\rangle) e^{-i\widetilde{E}_{0}t} + 2\sqrt{f_{0}}\text{Re}(\langle\wz^{\perp}|\wo\rangle) e^{-i\widetilde{E}_{1}t}] + \mathcal{O}(f)\\
\langle \alpha|\widetilde{U}(t)|\beta\rangle = \frac{1}{2}[e^{-i\widetilde{E}_{0}t} + ie^{-i\widetilde{E}_{1}t} + \sqrt{f_{0}}e^{-i\widetilde{E}_{1}t}(i\langle\wz^{\perp}|\wo\rangle + i\langle\wo|\wz^{\perp}\rangle)\\
+ \sqrt{f_{1}}e^{-i\widetilde{E}_{0}t}(i\langle\wz|\wo^{\perp}\rangle + \langle\wo^{\perp}|\wz\rangle) ] + \mathcal{O}(f)
\end{gather*}
And calculate the new probability:
\begin{gather*}
\widetilde{P}_{\alpha} = \frac{1}{2}[1 + \cos((\widetilde{E}_{1} - \widetilde{E}_{0})t)] + \mathcal{O}(\sqrt{f})\\
\widetilde{P}_{\beta} = \frac{1}{2}[1 + \sin((\widetilde{E}_{1} - \widetilde{E}_{0})t)] + \mathcal{O}(\sqrt{f})
\end{gather*}
The final value for phase difference is:
\begin{equation*}
\begin{aligned}
\delta\widetilde{\theta} &= \arctan\left[\frac{2\widetilde{P}_{\beta} - 1}{a\widetilde{P}_{\alpha} - 1}\right] \\
&= \arctan(\tan((\widetilde{E}_{1} - \widetilde{E}_{0})t) + \mathcal{O}(\sqrt{f}))\\
& = (\widetilde{E}_{1} - \widetilde{E}_{0})t + \frac{\mathcal{O}(\sqrt{f})}{1 + |\widetilde{E}_{1} - \widetilde{E}_{0}|^{2}t^{2}} + \cdots\\
&= \delta\theta + (\widetilde{E}_{1} - E_{1} - \widetilde{E}_{0} + E_{0})t + \mathcal{O}(\sqrt{f})
\end{aligned}
\end{equation*}
The second equation is only true when $\mathcal{O}(\sqrt{f}) \ll [1 + \cos((\widetilde{E}_{1} - \widetilde{E}_{0})t)]/2$.

We have proved that the trotter error in fidelity doesn't grow linearly with $L$. Thus, if $t$ is a constant instead of a small quantity, the effect of $\mathcal{O}(\sqrt{f})$ can be neglected. The final error in energy difference has order:
\begin{equation*}
|\delta\widetilde{\theta} - \delta\theta|/t = \mathcal{O}(|\widetilde{E} - E|) 
\end{equation*}

\end{document}